\documentstyle[preprint,prl,aps,12pt]{revtex}
\begin{document}
\draft
\title{Novel Quenched Disorder Fixed Point in a Two-Temperature Lattice Gas}
\author{B. Schmittmann and C. A. Laberge}
\address{Center for Stochastic Processes in Science and Engineering\\
and Department of Physics, Virginia Tech, Blacksburg, VA 24061-0435}
\date{December 9, 1996}
\maketitle

\begin{abstract}
We investigate the effects of quenched randomness on the universal
properties of a two-temperature lattice gas. The disorder modifies the
dynamical transition rates of the system in an anisotropic fashion, giving
rise to a new fixed point. We determine the associated scaling form of the
structure factor, quoting critical exponents to two-loop order in an
expansion around the upper critical dimension d$_c=7$. The close
relationship with another quenched disorder fixed point, discovered recently
in this model, is discussed.
\end{abstract}


The statistical mechanics of non-equilibrium phenomena remains poorly
understood, in spite of the ubiquity of such systems in nature. As a first
inroad towards the full complexity of such systems, simple model systems
driven into non-equilibrium steady states have attracted much interest over
the past decade \cite{review}. To prevent the systems from settling into the
usual Boltzmann equilibrium steady state, a uniform current (of particles or
energy) is maintained, e.g., by imposing a uniform drive, or by coupling the
system to two reservoirs at different temperature or chemical potential. The
prototype model is an Ising lattice gas, in the presence of a uniform
``electric'' field acting on the ``charged'' particles \cite{kls}. Motivated
by the physics of fast ionic conductors \cite{fic's}, the usual Ising
energetics is modified by the field which enhances (suppresses) the rate for
nearest-neighbor particle-hole exchanges along (against) a specific lattice
axis. The total number of particles is conserved, and periodic boundary
conditions are imposed to ensure a non-vanishing global current. The
resulting steady state distribution is non-Hamiltonian and violates detailed
balance. In spite of its simplicity, this model, and many of its variants,
exhibit surprising features like anomalous correlations at all temperatures
and different types of non-equilibrium universal behavior \cite{review}.

In the framework of renormalization group (RG) studies of these models, the
universality class of order-disorder transitions is determined by the
symmetries of the drive and the local order parameter alone, provided the
dynamics is short-ranged in space and time. For equilibrium systems it is
well known that spatial or temporal nonlocalities in the Hamiltonian, such
as sufficiently long-ranged interactions or quenched impurities, can have
profound consequences for critical behavior \cite{long-space,long-time}. On
the other hand, non-local terms appearing only in the dynamics but not in
the Hamiltonian have no effect on static universal properties provided
detailed balance holds. In contrast, if detailed balance is broken, the
steady state is determined by both the local energetics and the dynamical
rules, and nonlocalities in either are expected to play an important role.

So far, only a few studies have addressed this issue. The effects of
long-ranged random spin exchanges, added to the non-conserved Glauber
dynamics of the standard Ising model, have been investigated in some detail 
\cite{zr}. The resulting steady state, while Hamiltonian, is characterized
by effective long-range interactions between spins. Other studies have
focused on the effect of quenched disorder in the uniformly driven Ising
lattice gas \cite{bec+jan,lauritsen}. Here, due to the presence of a global
current, a subtle complication arises when attempting to compare field
theoretic results \cite{bec+jan} with Monte Carlo data \cite{lauritsen}. The
simulations are performed on a finite lattice with periodic boundary
conditions, so that particles typically see a given disorder configuration
repeatedly in the course of thermal averaging, thus inducing effective
correlations in the disorder average. In contrast, the analytic work rests
on the assumption of {\em uncorrelated} randomness.

Motivated by an interest in deepening our understanding of quenched disorder
in driven lattice gases, while recognizing the need to establish a clear
correspondence between Monte Carlo data and field theory results, a
non-equilibrium lattice gas {\em without }global current was studied in \cite
{sb}. In the absence of quenched disorder (henceforth called the pure case),
the field theory for this system captures the long-wavelength, long-time
behavior of two different microscopic models: a randomly driven Ising model 
\cite{sz} and a two-temperature lattice gas \cite{2T}. Adding a quenched
contribution to the drive in the randomly driven system, the authors showed
that the field theory of the pure model is significantly modified, giving
rise to a novel fixed point. The associated universal scaling behavior was
investigated to two-loop order and found to be distinct from both the pure
and the non-driven case.

In this Letter, we continue the analysis of quenched disorder in this
system. Modelling the disorder as a quenched random bias acting along
longitudinal or transverse bonds, we show that these two scenarios give rise
to two {\em distinct} fixed points. The former of these, subsequently
referred to as the ``longitudinal'' fixed point, was considered in \cite{sb}%
. Here, we investigate the other (``transverse'') fixed point. We begin by
summarizing the relevant properties of the pure system, followed by the
introduction of quenched randomness. Next, we discuss the resulting
effective Langevin equations for the longitudinal and transverse fixed
points. The universal scaling behavior associated with the latter is then
analyzed, using power counting techniques confirmed by an explicit $\epsilon 
$-expansion about the upper critical dimension of the theory. Explicit
results for critical exponents are computed to two-loop order. We conclude
with some comments.

In the absence of frozen disorder, our field theory describes the
coarse-grained properties of two different microscopic models. The first, a
randomly driven system \cite{sz}, corresponds to a (half-filled) lattice gas
of particles (up-spins) and holes (down-spins) on a fully periodic lattice,
interacting through an attractive nearest-neighbor Ising coupling and
coupled to a heat bath. As in the uniformly driven system, particle-hole
exchanges along a specific lattice axis are biased by a driving field. Here,
however, the {\em amplitude} $E$ of the drive switches sign and magnitude
randomly in space and time, i.e., a new value of $E$ is culled from a
symmetric distribution, $p(E)$, for each pair exchange along the field
direction. Exchanges along the transverse direction are controlled by
energetics alone. The symmetry of $p(E)$ ensures that there is no global
current; otherwise its detailed form is not important for critical behavior.
In its other version, a two-temperature model, we again consider a
half-filled Ising lattice gas with spin-exchange dynamics. The drive,
however, is now replaced by a second heat bath, such that spin pairs forming
a ``longitudinal'' bond exchange according to a temperature $T_{\Vert }$
while ``transverse'' bonds are controlled by a {\em different} temperature $%
T_{\bot }<T_{\Vert }$\cite{2T}. Both microscopic models obey the same set of
global symmetries, namely: translation invariance along the longitudinal and
the transverse directions, as well as rotation invariance in the $(d-1)$%
-dimensional transverse subspace. Moreover, both local order parameters are
scalar and exhibit the usual up-down symmetry of the Ising model. The other
common characteristic is that, as the (transverse) temperature is lowered,
long-wavelength fluctuations with purely transverse wave vector become soft
first, indicating that criticality sets in when the transverse diffusion
coefficient vanishes while the longitudinal one is still positive. A
Langevin equation for the local magnetization $\phi (x,t)$, consistent with
these features, is easily postulated, or derived from its counterpart for
the uniformly driven system by averaging with weight $p(E)$ \cite{sz}:

\begin{equation}
\lambda ^{-1}\partial _t\phi =(\tau _{\perp }-\nabla ^2)\nabla ^2\phi +\tau
_{\parallel }\partial ^2\phi +\frac g{3!}\nabla ^2\phi ^3-\partial j_{\Vert
}-\nabla j_{\bot }\text{ .}  \label{1}
\end{equation}
Here $\partial $ ($\nabla $) denotes the gradients in the parallel
(transverse) directions. $j_{\Vert }$ and $j_{\bot }$ are Gaussian white
noise terms which are distributed according to $\exp \{-\int d^dxdt[j_{\Vert
}{}^2/4n_{\Vert }+j_{\bot }{}^2/4n_{\bot }]\}$ and model the effect of
thermal noise after coarse-graining. The parallel and transverse diffusion
coefficients are denoted by $\tau _{\parallel }$ and $\tau _{\perp }$.
Criticality is associated with $\tau _{\perp }\rightarrow 0$ while $\tau
_{\parallel }$ $>0$. Since $\tau _{\parallel }$ does not vanish, $\tau
_{\parallel }\partial ^2\phi $ is the only relevant term involving $\partial
^2$. Otherwise, all coefficients here are functions of the microscopic
control parameters but their explicit form is not required. Standard field
theory methods \cite{ft} can be employed to compute the universal scaling
behavior of the theory \cite{sz}, in an expansion around the upper critical
dimension $d_c=3$. High precision Monte Carlo data for the two-temperature
model \cite{lpz} are consistent with the field theory predictions,
confirming that the two models do indeed belong to the same universality
class.

Next, we introduce frozen disorder into the {\em microscopic} dynamics of
the system, either in its random drive or two-temperature version, by adding
a quenched random bias $\Delta _\alpha (x)$ to the exchange rate of a
particle-hole pair located on a bond in the $\alpha $-direction, $\alpha
=\Vert ,\bot $. This type of randomness is easily implemented in a Monte
Carlo simulation, since it can be considered as a time-independent local
drive resident on each bond. In the Langevin equation (1), it leads to an
extra term, $\vec{\nabla}\cdot \vec{J}$, on the right hand side, with the
Ohmic current $J_\alpha $ given by $J_\alpha =c(\phi )\Delta _\alpha (x)$,
in analogy to the uniformly driven case \cite{review}. Here, $\vec{\nabla}$
is the full $d$-dimensional gradient, and $c(\phi )\propto 1+O(\phi ^2)$ is
the (density-dependent) conductivity. This form of $J_\alpha $ is also
easily derived from a microscopic hopping model, upon taking a naive
continuum limit and neglecting higher order derivatives. These, as well as
the $O(\phi ^2)$ correction to the conductivity, can be shown to be
irrelevant in the RG sense. The $\Delta _\alpha (x)$ are taken to be
Gaussian distributed with zero mean, $<\Delta _\alpha (x)>=0$ so that no
global current is induced, and $<\Delta _\alpha (x)\Delta _\beta (x^{\prime
})>=2\sigma _\alpha \delta _{\alpha \beta }\delta (x-x^{\prime })$. The
positive parameters $\sigma _{\bot }$ and $\sigma _{\Vert }$ control the
strength of the randomness. It is now apparent that the leading modification
of (1), apart from trivial redefinitions of its coefficients, is a quenched
random current, $-\partial \Delta _{\Vert }-\nabla \Delta _{\bot }$, on the
right hand side. To proceed, it is most convenient to introduce a
Martin-Siggia-Rose field $\tilde{\phi}(x,t)$ \cite{msr}, and to recast the
Langevin equation as a dynamic functional \cite{dyn_fun}. In this form, the
average over the $\Delta _\alpha (x)$ is easily performed, leading to an 
{\em effective }functional 
\begin{eqnarray}
&&{\cal J\,}[\phi ,\tilde{\phi}]=\int d^dx\int dt\lambda \tilde{\phi}\left\{
\lambda ^{-1}\dot{\phi}-\left[ (\tau _{\perp }-\nabla ^2)\nabla ^2\phi +\tau
_{\parallel }\partial ^2\phi +\frac g{3!}\nabla ^2\phi ^3\right] \right\}
\label{1} \\
&&+\int d^dx\int dt\lambda \int dt^{\prime }\lambda \tilde{\phi}(x,t)\left\{
(n_{\parallel }\partial ^2+n_{\perp }\nabla ^2)\lambda ^{-1}\delta
(t-t^{\prime })+(\sigma _{\parallel }\partial ^2+\sigma _{\perp }\nabla
^2)\right\} \tilde{\phi}(x,t^{\prime })\text{ .}  \nonumber
\end{eqnarray}
We emphasize that both the frozen disorder and the thermal noise generate
operators quadratic in $\tilde{\phi}$. In contrast to the thermal noise,
however, the quenched random contribution to the Langevin current leads to
an operator which is {\em nonlocal} in time. Correlation and response
functions can now be computed as functional averages with weight $\exp -%
{\cal J\,}[\phi ,\tilde{\phi}]$.

The next step in the RG treatment of (2) is a dimensional analysis. Since $%
\tau _{\bot }$ vanishes first, parallel momenta scale as $k_{\Vert }\propto
\mu ^2$, on a characteristic scale $\mu $ $\propto k_{\bot }$. Thus,
strongly anisotropic scaling \cite{review} is observed already at the tree
level. As a consequence, the ``transverse'' coefficient $n_{\perp \text{ }}$%
is more relevant than its ``longitudinal'' counterpart, $n_{\Vert }$, which
will therefore be dropped. Similarly, $\sigma _{\perp \text{ }}$, {\em if
non-vanishing}, is more relevant than $\sigma _{\Vert }$. A significant
advantage of our particular implementation of disorder, as a bias acting on
parallel or transverse bonds, now becomes apparent: it allows us to {\em tune%
} $\sigma _{\perp \text{ }}$or $\sigma _{\Vert }$ {\em separately} to $0$.
Here, we will choose $\sigma _{\perp \text{ }}\neq 0$, in contrast to Ref. 
\cite{sb}. We will return to the consequences of either choice in the
conclusions. Proceeding with the dimensional analysis, we can now neglect $%
\sigma _{\Vert }$ as well. Requiring that the remaining terms in (2) be
dimensionless, we find $\lambda t\propto \mu ^{-4}$, so that the thermal
noise becomes {\em irrelevant }and therefore negligeable compared to the
quadratic term associated with the frozen disorder. Furthermore, choosing
the scale of $\tilde{\phi}$ such that $\sigma _{\perp }\propto \mu ^0$, we
find $\phi \propto \mu ^{(d-5)/2}$ and $\tilde{\phi}\propto \mu ^{(d+7)/2}$,
implying that $g\propto \mu ^{7-d}$. The upper critical dimension follows as 
$d_c=7$, in contrast to the pure system where $d_c=3$. We also recall the
result $d_c=5$ for a disordered system with $\sigma _{\perp \text{ }}=0$ 
\cite{sb}. Thus, we may expect novel universal behavior for our theory. We
should remark that such dimensional shifts have also been found in the
uniformly driven system with frozen disorder \cite{bec+jan}.

We now consider the critical theory ($\tau _{\bot }=0$) with insertions of $%
\lambda \tau _{\perp }\tilde{\phi}\nabla ^2\phi $, to two-loop order in $%
\epsilon \equiv d_c-d$. A rescaling of $\tilde{\phi}$ allows us to set $%
\sigma _{\perp }=1$. The anisotropic forms of the bare correlation and
response propagators are easily obtained from the Gaussian parts of (2):
with $\Lambda (k)\equiv k_{\bot }^2(k_{\bot }^2+\tau _{\perp })+\tau _{\Vert
}k_{\Vert }^2$, we find $G_o(k,\omega )=[i\omega +\lambda \Lambda (k)]^{-1}$
for the response propagator, and $S_o(k,\omega )=4\pi \lambda ^2\delta
(\omega )|i\omega +\lambda \Lambda (k)|^{-2}$ for the correlator. The
latter, in particular, remains proportional to $\delta (\omega )$ at all
orders, due to the presence of the time-delocalized noise in (2). A simple
scale invariance of the theory, under $x_{\Vert }\rightarrow \gamma x_{\Vert
}$ at constant $x_{\bot }$ with arbitrary scale factor $\gamma $, identifies
the effective dimensionless expansion parameter of the theory as $u\equiv
\mu ^{-\epsilon }\tau _{\Vert }^{-1/2}g$. Denoting one-particle irreducible
vertex functions with $n$ ($\tilde{n}$) external $\phi $- ($\tilde{\phi}$-)
legs and $m$ insertions by $\Gamma _{\tilde{n}n}^{(m)}$, we find that only $%
\Gamma _{11}^{(0)}$, $\Gamma _{13}^{(0)}$, and $\Gamma _{11}^{(1)}$ are
primitively divergent. Thus, only the quantities $\phi $, $u$, and $\tau
_{\perp }$ must be renormalized to render the theory finite. The coupling
constant $u$ flows towards an infrared stable fixed point $u^{*}$, while the
renormalizations of $\phi $ and $\tau _{\perp }$ determine the only two {\em %
independent} critical exponents, $\eta $ and $\nu $. Following standard
methods \cite{ft}, the renormalization group equations at the fixed point,
combined with dimensional analysis and scale invariance, predict the full
asymptotic scaling behavior of the vertex functions $\Gamma _{\tilde{n}%
n}^{(m)}$. Instead of quoting the complete expressions here, we specialize
to the steady-state structure factor, which is easily measured in
simulations: 
\begin{equation}
S(k,t;\tau _{\bot })=\mu ^{-6+\eta }S(k_{\parallel }/\mu ^{1+\Delta
},k_{\perp }/\mu ,t\mu ^z;\tau _{\perp }/\mu ^{1/\nu })  \label{3}
\end{equation}
Some comments are in order here. First, the prefactor $\mu ^{-6+\eta }$ may
appear somewhat surprising, compared to the more familiar $\mu ^{-2+\eta }$.
However, its origin is easily traced back to the tree level result for the
structure factor, $S_o(k,\omega )$, where the $\delta (\omega )$ is
responsible for changing the familiar power of $2$ to the more unusual $6$.
Second, the strong anisotropy exponent $\Delta $ \cite{review} and the
dynamic exponent $z$ are related to the two independent exponents $\eta $
and $\nu $ via scaling laws: $\Delta =1-\eta /4$ and $z=4-\eta /2$. The
latter differs from the familiar $z=4-\eta $ (for the conserved Ising
model), again by virtue of the quenched disorder which modifies the usual
relationships between the Wilson functions for $\phi $, $\tilde{\phi}$, and $%
\lambda $. The details of the two-loop calculation are too technical to be
reported here, but the stability of the fixed point is established and
explicit results for the critical exponents can be computed. We find $\eta
\equiv \frac{88}{2187}\epsilon ^2+O(\epsilon ^3)$ and $\nu =\frac 12+\frac 
\epsilon {12}+\frac{\epsilon ^2}{324}\left( \frac{1081}{81}+\ln \frac 43%
+6\ln 4\right) +O\left( \epsilon ^3\right) $, which shows clearly that our
theory falls into a new universality class.

Other scaling laws determine the remaining anisotropic exponents \cite
{review}, such as the two different correlation length indices $\nu _{\perp
}=\nu $ and $\nu _{\parallel }=\nu (1+\Delta )$, or the two different
dynamic exponents $z_{\perp }=z$ and $z$ $_{\parallel }=z/(1+\Delta )$. Due
to the anisotropy, care must also be taken when defining exponents that
characterize the power-law decays of the critical structure factor, in real
and momentum space. Simple scaling laws relate these indices to $\eta $ and $%
\Delta $ \cite{review}. Finally, the order parameter exponent $\beta =\frac 1%
2\nu (d-6+\eta +\Delta )$ follows from the scaling form of the equation of
state \cite{ft}.

In summary, we have analyzed the critical behavior of a two-temperature
stochastic lattice gas with conserved dynamics, in the presence of a
quenched random bias acting on bonds in both the longitudinal and transverse
subspace. The bias generates a time-delocalized Langevin current which
modifies the usual power counting so dramatically that the upper critical
dimension is shifted to $d_c=7$. The RG analysis yields the asymptotic
scaling forms of correlation and response functions; and critical exponents,
associated with a new universality class, have been computed to $O(\epsilon
^2)$ in an expansion about $d_c$. If, in contrast, the frozen disorder
affects {\em longitudinal }bonds only, then $\sigma _{\bot }$ $=0$ in Eqn
(2). In this case, the dimensional analysis has to be revisited, resulting
in $d_c=5$. Moreover, an additional relevant operator $-\kappa \int d^dx\int
dt\int dt^{\prime }\lambda ^2\tilde{\phi}(x,t)\nabla ^4\tilde{\phi}%
(x,t^{\prime })$ is {\em generated} under the RG and must be included in the
dynamic functional. Considering the interplay between the two couplings $%
\sigma _{\Vert }$ and $\kappa $, the former is found to be irrelevant while
the latter gives rise to a new (``longitudinal'') fixed point which is
infrared stable near $d=5$ \cite{sb}. It is noteworthy \cite{sb} that this
fixed point also captures the effect of quenched disorder resident in the
nearest-neighbor {\em hopping rates}, modelled, e.g., by a Gaussian
distribution of potential barriers between sites. Eqn (2), including the
operators associated with $\sigma _{\Vert }$, $\sigma _{\bot }$ and $\kappa $
thus provides a unified field theory for several different types of quenched
disorder. Work is in progress to clarify the intriguing connections between
different microscopic implementations of such disorder and their surprising
degree of universality \cite{sz1}.\newpage\ 

Acknowledgements:

We wish to thank K.E.\ Bassler, V. Becker, H.K. Janssen, K. Oerding and
R.K.P. Zia for many helpful discussions. We are particularly grateful to V.
Becker and H.K. Janssen for communicating the results of Ref. \cite{bec+jan}
to us prior to publication. This work is supported in part by the National
Science Foundation through DMR-9419393.

\end{document}